\documentstyle[prd,aps,floats]{revtex}
\begin{document}
\draft

%
%
\input epsf \renewcommand{\topfraction}{0.8} 
\twocolumn[\hsize\textwidth\columnwidth\hsize\csname 
@twocolumnfalse\endcsname

\title{Extra Dimensions and Inflation} 
\author{Anupam Mazumdar}
\address{Astrophysics Group, Blackett Laboratory, Imperial College London,
 SW7 2BZ, ~~~U.~K.}
\date{\today} 
\maketitle
\begin{abstract}
It has been proposed that, without invoking supersymmetry, it is 
possible to solve the hierarchy problem provided the fundamental scale 
in the higher dimensional theory is at a much lower scale than the 
Planck scale. In this paper we consider a toy model where we allow 
$4+d$ dimensions to evolve from a region determined by the new 
fundamental scale, which is in our case is electro-weak scale. We
further investigate whether it is possible to inflate not only the $3$ 
uncompactified dimensions but also the extra dimensions. We require
around $70$ e-foldings to stabilize the size of two extra dimensions from
$M_{\rm EW}^{-1}$ to the required $1$ mm size. This can be easily 
achieved during inflation, but once inflation ends their
evolution is governed by the dynamical history of the universe in a 
similar way that of Brans-Dicke field in generalized Einstein theories.
We also show that we achieve the right level of density contrast 
without invoking any specific potential. The density contrast 
depends on the number of extra dimensions and upon the amplitude of
the potential. 
\end{abstract}

\pacs{PACS numbers: 98.80.Cq \hspace*{1.3cm} Imperial preprint Imperial-AST 
99/2-2, hep-ph/yymmmnn}

\vskip2pc]


\section{Introduction}
The longstanding hierarchy problem, that Higgs fields have 
mass $m_H \sim 1$ TeV and not the Planck mass $M_{\rm p}$ although through 
the loop corrections one expects the correction should be of the order 
of the cut off which is $M_{\rm p} \sim 10^{18}$ GeV, is solved by 
supersymmetry 
by cancelling the loop contribution from fermions and bosons 
to render such low mass for the Higgs fields. However, this problem can 
be solved without invoking supersymmetry, as shown in Ref.{\cite{nima}}. If 
one begins with the higher dimensional theory as a fundamental theory 
and also considers the scale of gravity to be gauge unification scale 
instead of $M_{\rm p}$ in $4+d$ dimensions, where $d$ represents the 
number of extra dimensions, then it is possible to recover $M_{\rm p}$  
in the $3+1$ dimensional wall where the standard model fields are distributed.
Such a model seriously undermines the structure formation if the evolution
of extra dimensions are not taken into account \cite{lyth}. The 
importance of extra dimensions in dynamical evolution
has been discussed in various papers
\cite{linde,gia}. Here we follow 
a similar approach, allowing $4+d$ dimensions to evolve from the 
electro-weak scale. We consider the matter lagrangian in $4+d$ dimensions
and then we reduce the lagrangian to $4$ dimensions by taking a
simple ansatz for the extra dimensions to be compactified on $d$ dimensional
torus.
At no stage of our calculation do we consider any specific potential
for the matter lagrangian. All we assume is that the potential has a
global minimum. By considering this we show that it is possible to inflate
the $3$ spatial uncompactified dimensions exponentially in the string frame and
the extra compactified dimensions to grow from the $M_{\rm EW}^{-1}$ scale to
$1$mm size required by the two extra dimensions. To evolve the extra
dimensions we require a minimum $70$ e-foldings of inflation. Once 
inflation ends the evolution of the extra dimensions is governed by the
dynamical history of the universe similar to the Brans-Dicke field.  
We also consider the density perturbation created during the inflationary
phase in the Einstein frame and conclude that it is easy to 
produce the required density
contrast by constraining the amplitude of the potential from the COBE
data. It is worth mentioning that our density perturbation results 
depend on the number of extra dimensions and also upon the amplitude 
of the potential. 


\section{Equations of Motion}
We consider our lagrangian in $4+d$ dimensions:
\begin{eqnarray}
\label{lag}
S = \int d^{4+d}X \sqrt{-G}\left[ \frac{1}{2\hat{\kappa}^2_{4+d}}
\hat R + L_{\rm inf}\right]\,,
\end{eqnarray}
where $G$ and $\hat{\kappa}^2_{4+d}$ are the $4+d$ dimensional metric and 
gravitational constant. $L_{\rm inf}$ is the lagrangian for the scalar field  
with potential in $4+d$ dimensions. By taking an ansatz that the line-element
in $4+d$ dimensions takes the following form:
\begin{eqnarray}
d\hat s^2=dt^2 -a(t)^2g_{\mu\nu}dx^{\mu}dx^{\nu} -b(t)^2g_{ij}dx^{i}dx^{j}\,,
\end{eqnarray}
where $a$ is the scale factor of the $3$ dimensions and $b$ is the scale
factor of the extra dimensions. The geometry of $g_{ij}$ is assumed to be 
torus with a unit volume.
It is possible to reduce 
Eq.(\ref{lag}) to a $4$ dimensional lagrangian. In $4$ dimensions the 
lagrangian mimics the lagrangian of Jordan--Brans--Dicke theory where there
is an extra field related to the size of the extra dimensions coupled to 
the Ricci scalar in $4$ dimensions. For details, see \cite{kolb} and 
\cite{berkin}.
The reduced lagrangian in $4$ dimensions in the string frame is:
\pagebreak
\begin{eqnarray}
\label{redflag}
S=\int d^4x\sqrt{-g}\left[-\Phi R -\frac{d-1}{d}\frac{(\nabla \Phi )^2}{\Phi}\,
+ {2\kappa ^2 \Phi}\frac{1}{2}(\nabla \chi )^2 \,
\right. \nonumber \\
\left. - 2\kappa^2 \Phi V(\chi )\right]\,,
\end{eqnarray}
where $\kappa $ and $g$ are the $4$ dimensional gravitational constant and the
metric. $\kappa $ is related to $4+d$
dimensional gravitational constant by
\begin{eqnarray}
\label{ka1}
\kappa ^2 = \frac{\hat \kappa^2_{4+d}}{2^d b_0^d \pi}\,.
\end{eqnarray}
$\chi $ and $V(\chi )$ are the corresponding inflaton and its potential
in the reduced $4$ dimensions and $\Phi $ is defined by the scale of 
compactification.
\begin{eqnarray}
\label{rew0}
\Phi \equiv \frac{1}{2\kappa^2}\left[\frac{b}{b_0}\right ]^{d}\,,
\end{eqnarray}
where $b_0$ is the scale factor of the extra dimensions at present.
we must mention that the $4+d$ dimensional gravitational constant can be 
recast according to the demand of the newly proposed scale:
\begin{eqnarray}
\label{hatk}
\hat \kappa ^2_{4+d}=\frac{8\pi}{M_{\rm {EW}}^{2+d}}\,.
\end{eqnarray}
In the denominator of Eq.(\ref{hatk}), instead of $M_{\rm p}$
we have a new fundamental scale
determined by the electro-weak unification scale $M_{\rm EW}$. Hence, the $4$
dimensional gravitational constant can be expressed by:
\begin{eqnarray}
\kappa ^2 = \frac{8\pi}{2^d b_0^d \pi M_{\rm{EW}}^{(2+d)}}\,.
\end{eqnarray}
For our purpose we take $b_0$ to be $1$mm for $2$ extra
dimensions if $M_{\rm{EW}}\sim 1$TeV as proposed in Ref.{\cite{nima}}.
It is worth mentioning at this point that the Planck mass in $4$ dimensions
is determined by:
\begin{eqnarray}
\label{pl}
M_{\rm p}^2 =2^d b_0^d \pi M_{\rm EW}^{(2+d)}\,.
\end{eqnarray}

We may now write down the equations of motion corresponding to 
Eq.(\ref{redflag}) in the string
frame and later on we shall analyze in the Einstein frame also. In a spatially
flat Robertson-Walker universe:
\begin{eqnarray}
\label{eq1}
H^2 +H\frac{\dot\Phi}{\Phi}&=&-\frac{1}{6}\left[\frac{\dot\Phi}{\Phi}\right]^2
+\frac{\kappa^2}{3}\left[\frac{1}{2}\dot\chi^2 +V(\chi)\right]\,, \\
\label{eq2}
\ddot\Phi +3H\dot\Phi &=&\frac{2\kappa^2 \Phi }{1+2/d}\left[V(\chi)-\dot\chi^2
\right]\,,\\ 
\label{eq3}
\ddot\chi +3H\dot\chi &=& -V'(\chi)\,.
\end{eqnarray}
An overdot denotes derivative with respect to time and prime with
respect to the inflaton field $\chi $. $H$ is the Hubble constant of
the universe, $H=\dot a/a$. Irrespective of the form of the
potential $V(\chi )$, if inflation occurs then following the slow-roll
approximation \footnote{Appropriate slow-rollover approximations are
$|\frac{\dot\Phi}{\Phi}| \ll H $, $|\ddot \Phi | \ll 3H\dot \Phi$, 
$|\dot \chi^2 | \ll V(\chi)$, and $|\ddot\chi | \ll 3H\dot \chi $.}
one can reduce the above Eqs. (\ref{eq1}--\ref{eq3}).
\begin{eqnarray}
\label{simp1}
H^2 &\approx & \frac{\kappa^2}{3}V(\chi)\,, \\
\label{simp2}
3H\frac{\dot\Phi}{\Phi} &\approx & \frac{2\kappa^2 V(\chi)}{1+2/d}\,, \\
\label{simp3}
3H\dot \chi &\approx &  -V'(\chi)\,.
\end{eqnarray}
Solving Eqs.(\ref{simp1}-\ref{simp2}) we get the following solutions 
\cite{berkin}:
\begin{eqnarray}
\label{rel1}
\frac{\dot\Phi}{\Phi}= d\frac{\dot b}{b} & = &\frac{2d}{2+d}
\frac{\dot a}{a}\,, \\
\label{rel2}
a &\propto & e^{Ht}\,.
\end{eqnarray}
Where we have used Eq.(\ref{rew0}) in Eq.(\ref{rel1}) to derive the 
most important relation between the scale factors which enable us to 
stabilize the extra dimensions. Eq.(\ref{rel2}) clearly shows that the 
scale factor in the uncompactified $3$ dimensions grow exponentially fast
provided the potential $V(\chi )$ is sufficiently flat and treated almost
as a constant during the inflationary phase.
This also suggests that the extra dimensions are also inflating but their
rate of expansion solely depends on the number of extra dimensions governed
by Eq.(\ref{rel1}).

Now let us analyze the Hubble parameter during inflation. Before that we
must note that the effective potential in $4$ dimensions can at most take
the value $b_0^d M_{\rm{EW}}^{4+d}$ and not just $M_{\rm{EW}}^4$
\footnote{In general, the potential $V(\chi )$ would be of the order of
$M_{\rm EW}^4$ had we neglected the dynamics of extra dimensions, but
a careful analysis suggests that on dimensional grounds
$V(\chi)\approx M^4 \equiv b_0^d M_{\rm EW}^{4+d}$; for details,
see Eqs. (3.4) and (3.5) in \cite{kolb}. It is extremely important to note
that in $4$ dimensions the effective potential should have significant
energy contribution from the extra dimensions. This can be vividly seen
by arguing that if we do not allow the extra dimensions to evolve, $b
\approx b_0$, we get the energy density in $4$ dimensions, $V(\chi )\approx
M_{\rm EW}^4$.}. Here we
must mention that had we introduced the inflaton in the $4$ dimensional
lagrangian we would have got the latter bound on $V(\chi)$.
Here we get the following relation:
\begin{eqnarray}
\label{hubb}
H^2 \approx \frac{8}{3\cdot 2^{d}b_0^d M_{\rm EW}^{(2+d)}}M_{\rm EW}^{(4+d)}
b_0^d \sim \frac{8 M_{\rm EW}^2}{3\cdot 2^d} \,.
\end{eqnarray}
This is an important result as it suggests that the Hubble parameter can 
roughly take the electro-weak scale in the string frame. Now it is worth
questioning the fate of the extra diemnsions. In fact our simple analysis 
shows that the extra diemsnions will also grow exponentially fast but they
will never be able to overcome the $1$ mm size as expected for two extra 
dimensions, because there is a vital difference between the rate of expansions
of the extra dimensions and the observable uncompactified universe.

Once inflation ends in this model the $\chi $ field oscillates at the bottom
of the potential and it can be shown with the help of Eq.(\ref{eq2}) that on
average the right hand side of the Eq.(\ref{eq2}) vanishes, leading to 
a late time attractor solution for $\Phi = \it{Constant}$, the constant 
value of $\Phi$ is 
fixed by the present value of the Newton's constant and thus stabilizes the 
size of the extra dimensions which should be of the order
of $1$mm. for two extra diemensions. This leads to a novel dynamical method
to stabilize the extra dimensions. Hence, in our model the present value of
the Newton's constant is fixed right after the end of inflation and during 
the subsequent evolution of the universe, the value of the Newton's constant
remain unchanged.

\section{Density Perturbation}
We carry out the density perturbation calculation in the Einstein frame. In
fact the string frame is related to the Einstein frame through the 
conformal transformation $g_{\mu\nu}=(2\kappa^2 \Phi)^{-1}\tilde g_{\mu\nu}$
and in terms of field:
\begin{eqnarray}
\label{confo} 
\kappa \phi = \left[\frac{d+2}{2d}\right]^{1/2}ln\left[2\kappa^2 \Phi
\right]\,,
\end{eqnarray}
where $\phi$ is the field in the Einstein frame corresponding to the field
$\Phi$ in the string frame. The reduced $4$ dimensional lagrangian in the 
Einstein frame is:
\begin{eqnarray}
\label{redein}
S=\int d^4x\sqrt{-\tilde g}\left[-\frac{1}{2\kappa^2}\tilde R+\frac{1}{2}
(\tilde \nabla \phi )^2
+\frac{1}{2}(\tilde \nabla \chi )^2 \,
\right. \nonumber \\
\left. - e^{-\beta \kappa \phi }V(\chi)  )\right]\,,
\end{eqnarray} 
where $\beta $ in our case is defined by the following relation:
\begin{eqnarray}
\label{ba}
\beta =\left[\frac{2d}{d+2}\right]^{1/2}\,.
\end{eqnarray}
The advantage of the conformal transformation is that it allows the
density spectrum and the reheating temprature to be derived using the well 
known results from the Einstein gravity \cite{kalara} and \cite{jim}.
It is easy to derive the
slow-roll equations of motion in the Einstein frame by neglecting the kinetic 
and the double time derivative terms.
\begin{eqnarray}
\label{slowr1}
3\tilde H\dot \phi &\approx &\beta \kappa e^{-\beta \kappa \phi} V(\chi) \,, \\
\label{slowr2}
3\tilde H\dot \chi &\approx & -e^{-\beta \kappa \phi} V^{\prime}(\chi) \,, \\
\label{slowr3}
\tilde H^2 &\approx & \frac{\kappa^2}{3}e^{-\beta \kappa \phi}V(\chi )\,.
\end{eqnarray} 
Here the dot is differentiation with respect to the time in the Einstein frame.
It is worth mentioning that the Hubble parameter in the Einstein frame will
not be of the order of electro-weak scale but it is modified by the conformal 
factor. In fact it is very easy to show with the help of Eq.(\ref{rew0})
and Eq.(\ref{confo}) that the Hubble parameter in the Einstein frame is 
close to the Planck scale in $4$ dimensions. This ensures that the number
of e-foldings in both the frames are roughly going to be equal
\footnote{
Number of e-foldings defined in the string frame: $N=\int dt H$, and in the
Einstein frame $\tilde N =\int d\tilde t \tilde H$. Now, from the 
conformal transformation $H=\Omega \tilde H -\dot \Omega $. In our case
$\Omega^2 = 2\kappa ^2 \Phi $. If $|\frac{\dot \Omega}{\Omega \tilde H}| \ll 1$,
then, $H \approx \Omega \tilde H$ and subsequently we find $N \approx \tilde N$.
In our case $|\frac{\dot \Omega}{\Omega \tilde H}| =\frac{d}{d+2}\frac{H}{
\tilde H} \approx \frac{d}{d+2} \frac{M_{\rm EW}}{M_{\rm p}} \approx 10^{-17}
\ll 1$.
This ensures that the observable quantities in both the frames are 
equivalent and
can be translated from one frame to the other.}.
To carry out the density perturbation calculation we simply 
assume that one of the 
scalar fields dominates the other and we also use the above equations 
Eq.(\ref{slowr1}-\ref{slowr3}) to derive the form of the density contrast.
In the Einstein frame when  
$\dot \phi \ll \dot \chi $: 
\begin{eqnarray}
\label{den1}
\frac{\delta \rho}{\rho} = \frac{\tilde H^2}{2\pi\dot \chi}=\frac{\kappa^3}
{2\pi \sqrt{3}}\frac{V_h^{3/2}}
{V^{\prime}_h}
\left[\frac{b_h}{b_0}\right]^{-d/2}\,, 
\end{eqnarray}
and, when $\dot \chi \ll \dot \phi $:
\begin{eqnarray}
\label{den2}
\frac{\delta \rho}{\rho}=\frac{\tilde H^2}{2\pi \dot\phi}=\frac{\kappa^2}
{2\pi \sqrt{3}\beta}\sqrt{V_h}
\left[\frac{b_h}{b_0}\right]^{-d/2}\,,
\end{eqnarray}
where subscript $h$ denotes that the corresponding quantities are
evaluated at the time of horizon crossing. We assume that the amplitude
of the inflaton potential is determined by a dimensionless parameter
$\alpha$, such that
$V(\chi)\propto \alpha b_0^d M_{\rm EW}^{4+d}$. The detailed form of the 
potential
is not required for our analysis. Further assuming that the compactification
scale during horizon crossing is very close to the electro-weak scale, 
one can easily deduce the density contrast for two different cases. 
Eq.(\ref{den1}) reduces to a simpler form:
\footnote{
$V^{\prime}(\chi)
\sim V(\chi)/\chi$, and from dimensional analysis 
$\chi \approx M \equiv b_0^{d/2}M_{\rm EW}^{1+d/2}$,
we get $V^{\prime}(\chi) \approx b_0^{d/2}M_{\rm EW}^{3+d/2}$. For details,
see Eqs.(3.5) and (3.7) in \cite{kolb}.}

\begin{eqnarray}
\label{den3}
\frac{\delta \rho}{\rho}=\frac{(8\alpha)^{3/2}}{2^{3d/2+1}\sqrt{3}\pi}\,, 
\end{eqnarray}
and Eq.(\ref{den2}) reduces to:
\begin{eqnarray}
\label{den4}
\frac{\delta \rho}{\rho}= \frac{8\alpha^{1/2}}{2^{d+1}\sqrt{3}\pi \left[
\frac{2d}{d+2}\right]^{1/2}} \,.
\end{eqnarray}
The value of the amplitude of 
the potential is to be 
adjusted to give the right level of perturbations when our present
Hubble scale crossed outside the horizon during inflation. The 
perturbations observed by COBE satellite require $\delta \rho/\rho \approx
2 \times 10^{-5}$; this determines the value of $\alpha$ in our case; $\alpha
=1.1\times 10^{-3}$ when $\dot \phi \ll\dot \chi$, and $\alpha =1.2\times
10^{-10}$ while taking the opposite limit for $d=2$. 
For a quadratic potential in $4$
dimensions ( a quadratic potential enjoys the dimensional consistency
in $4$ dimensions as well as in higher dimensions )
the mass of the inflaton requires $\alpha ^{1/2}m_{\chi}\sim 0.03
M_{\rm EW}$ when $\dot \phi \ll \dot \chi$, suggest that the inflaton 
mass is very close to the electro-weak scale. At this point it is 
worth mentioning that the density contrast result is quite sensitive 
to the number of extra dimensions and the scale invariance of the density
perturbation is already treated in the context of "soft inflation", see
\cite{berkin}. 

\section{Stabilizing Extra Dimensions}
Next we consider the issue of stabilizing the size of the extra dimensions.
As we have already shown that the $3$ uncompactified spatial dimensions 
grow exponentially fast and through Eq.(\ref{rel1}) it is clear that
the scale factor for the extra dimensions also grow exponentially 
fast. Fortunately the rate of expansions for $a$ and $b$ are different
and this is solely responsible to stabilize the size of the extra dimensions 
soon after the end of inflation. We argue that 
as soon as inflation stops, the extra dimensions stop growing and stabilize
close to the observed value $b_0$, similar to the Brans-Dicke field which
saturates as soon as inflation ends. Solving Eq.(\ref{rel1}) 
we can estimate the evolution of $b$ during inflation.
\begin{eqnarray}
\label{gro}
\frac{b_f}{b_i} = \left[\frac{a_f}{a_i}\right]^{\frac{2}{2+d}} \,,
\end{eqnarray}
where the subscripts stands for final and initial values. Now we 
observe that for the extra dimensions to grow from $1$ TeV scale 
which corresponds to $10^{-16}$cm to $1$mm size, we require at least
$35$ e-foldings of inflation in the extra dimensions. Hence, the
left hand side of the Eq.(\ref{gro}) will be exactly $e^{35}$. If
we assume that there are only two extra dimensions then the right hand
side of the Eq.(\ref{gro}) suggests that the scale factor in the 
uncompactified dimensions will grow $e^{70}$, which is a reasonable 
number of e-foldings to solve the usual cosmological problems in the 
uncompactified dimensions. Here we must say that the analysis is 
independent of the details of the model and hence we do not require 
any fine tuning in the parameters of the inflaton potential.
It is important to note that for $d >2$, one requires more than
$70$ e-foldings of inflation in the uncompactified dimensions 
to stabilize the extra dimensions. In fact 
further evolution of the extra dimensions 
depends on the dynamical evolution of the universe, similarly to the 
Brans-Dicke case, this claim is further validated by inspecting
Eq.(\ref{eq2}), that, when the $\chi $ field oscillates at the bottom 
of the potential
at the end of inflation and on average cancelling the potential term with
the kinetic term, the dynamical stabilization of the $\Phi $ field is 
achieved automatically. In fact this is a novel way to stabilize the extra
dimensions where we need not require to invoke any potential to fix the 
value of the Newton's constant.

\section{Conclusions}
In this paper we have discussed that inflation at the TeV unification scale
can occur in a usual sense with the Hubble constant $H^{-1} \approx
M_{\rm EW}^{-1}$ in the string frame. The only fundamental energy density
is $M_{\rm EW}^{4+d}$. the energy density in $4$ dimensions is largely
contributed by the energy density from the extra dimensions. Our results 
are quite generic and we hardly require
any fine tuning in our potential. Density perturbations in this model
depends on the number of extra dimensions and on the amplitude of the
potential. We must say that our density perturbation calculation is 
valid only in the Einstein frame but we have also shown that 
it is adequate to discuss the perturbation analysis in the Einstein frame 
because string frame is conformally related to the Einstein frame. We 
show that indeed it is possible to stabilize the size of the extra dimensions.
For two extra dimensions to stabilize at $1$mm size one needs to invoke
roughly $70$ e-foldings of expansion and once 
inflation ends the fate of the extra dimensions
is determined by the dynamical history of the
universe, similar to the case of the Brans-Dicke field. We must mention that
here we need not invoke any apriori potential to fix the dilaton, rather 
the dilaton is fixed by the dynamical stabilization to give rise to the
value of the Newton's constant depending on the present size of the 
extra dimensions.


\section*{Acknowledgments}

A.M. is supported by the Inlaks foundation and an ORS award.

\end{document}